# Educational Data Mining and Learning Analytics - Educational Assistance for Teaching and Learning

Ganesan Kavitha[1], Lawrance Raj[2]

[1]Computer Science & Engineering Department, Jubail University College
Jubail Industrial City, Eastern Province, 31961, Saudi Arabia
[2]Department of MCA, Ayya Nadar Janaki Ammal College
Sivakasi, Tamil Name, 626123, India

***Abstract:*** *Teaching and Learning process of an educational institution needs to be monitored and effectively analysed for enhancement. Teaching and Learning is a vital element for an educational institution. It is also one of the criteria set by majority of the Accreditation Agencies around the world. Learning analytics and Educational Data Mining are relatively new. Learning analytics refers to the collection of large volume of data about students in an educational setting and to analyse the data to predict the students' future performance and identify risk. Educational Data Mining (EDM) is develops methods to analyse the data produced by the students in educational settings and these methods helps to understand the students and the setting where they learn. Aim of this research is to collect large collection of data on students' performance in their assessment to discover the students at risk of failing the final exam. This analysis will help to understand how the students are progressing. The proposed research aimed to utilize the result of the analysis to identify the students at risk and provide recommendations for improvement. The proposed research aimed to collect and analyse the result of the assessment at the course level to enhance the teaching and learning process. The research aimed to discuss two feature selection techniques namely information gain and gain ratio and adopted to use gain ratio as the feature selection technique.*

**Keywords:** *Educational Data Mining, Learning Analytics, Accreditation, Retention, Assessment, and Teaching and Learning.*

## 1. INTRODUCTION

### 1.1 What is Educational Data Mining (EDM)?

Educational data mining tends to focus on developing new tools for discovering patterns in data. These patterns are generally about the micro conceptsinvolved in learning [6]. According to International Educational Data Mining Society [4], EDM is an emerging discipline, concerned with developing methods for exploring the unique and increasingly large-scale data that come from educational settings, and using those methods to better understand students.

### 1.2 What is Learning Analytics (LA)?

Tanya [9] says in her report "learning analytics is an emerging field in which sophisticated analytic tools are used to improve learning and education. It draws from, and is closed tied to, a series of other fields of study including business intelligence, web analytics, academic analytics, educational data mining, and action analytics.

According to Mark and David [7], some examples of the use of analytics for learning and teaching are to:
- Identify students at risk so as to provide positive interventions designed to improve retention.
- Provide recommendations to students in relation to reading material and learning activities.
- Detect the need for, and measure the results of, pedagogic improvements.
- Tailor course offerings.
- Identify teachers who are performing well, and teachers who need assistance with teaching methods.
- Assist in the student recruitment process.

This paper aims to help the students' learning process by analysing the students' assessments to:
- Help the educators / teachers to identify the students at risk, and also to identify the patterns.
- Help the students to further improve their learning process.

This research will also help the students' learning process and the instructors' teaching process as teaching and learning are the vital element of an educational institution. This research aims to develop an application to read the outcome of the students' various assessments to find out the list of students who are at risk status and also to analyse each assessment to understand the students' progress in a particular course.

## 2. BACKGROUND AND LITERATURE REVIEW

Ramaswami and Bhaskaran [5], analysed six feature selection techniques such as Correlation-based Attribute Evaluation, Chi-Square Attribute Evaluation, Gain-Ratio Attribute Evaluation, Information Gain Attribute Evaluation, Relief Attribute Evaluation and Symmetrical Uncertainty Attribute Evaluation. Based on ROC Values and F1-Measure, they arrived at the conclusion that Information Gain technique is the best feature selection technique.

Asha and her team [1], used C4.5 decision tree algorithm which is the successor of ID3 algorithm. ID3 algorithm using information gain as feature selection technique whereas C4.5 uses Gain Ratio as feature selection technique. The attribute with the highest gain ratio is chosen as the spitting attribute. The non-leaf node of the decision tree generated are the most relevant attributes.

Baradwaj and Saurabh Pal [2] analysed the students' performance in a course to predict their performance in the end semester examination. In their research, they applied Gain Ratio to as feature selection technique. The concluded that various attributes like Previous Semester Marks, Class Test Grade, Seminar Performance, Assignment, General Proficiency, Attendance, Lab Work plays an important role in predicting students' performance in the end semester examination.

## 3. APPLICATION FOR PROVIDING EDUCATIONAL ASSISTANCE

### 3.1 Data Preparation
The training set used in this study was received from Jubail University College, Kingdom of Saudi Arabia. A sample of 20 students' records were selected for this research.

### 3.2 Data Selection

In an Educational System, the success of a student is determined by the coursework and final examination. In Jubail University College, the coursework is carried out throughout the semester as quizzes, mid-term examination, and assignments. The final examination is conducted at the end of the semester. In order to pass a course, the student has to score the minimum marks in the total of the coursework and final examination. The student who could not able to get the minimum score has to re-take the course in the following semesters. Students Related Attributes are given in Table 1.

**TABLE 1. STUDENTS RELATED ATTRIBUTES**

| Attributes | Standard Norms |
|---|---|
| Quiz 1/Quiz 2 | {Pass>=6, Fail<6} |
| Assignment 1 | {Pass>=4.8,Fail<4.8}for course with practical<br>{Pass>=6, Fail<6} for course without practical |
| Assignment 2 | {Pass>=7.2, Fail<7.2} for course with practical<br>{Pass>=6, Fail<6} for course without practical |
| Mid Term Exam | {Pass>=12, Fail<12} |

The author has developed a small application in Java to analyse the students' performance of a particular course. The outcome of the students assessments for a particular course is provided to the application as an Excel file. The coursework is for 60 marks and the final is for 40 marks and together for 100 marks. The students' needs to get the minimum marks, which is 60, to pass a course. Courses which have practical lessons will have the assessments for theory and practical separately such as Assignment LT, Assignment LB, Mid Term Theory, and Mid Term Practical. Courses which do not have practical lessons will have the consolidated assessments like Assignment 1, Assignment 2, Mid Term, and Final. The Quiz 1 and Quiz 2 are common for all types of courses.

All the assessments, excluding Final, is consolidated for 60 marks. Before the students appear for the final exam, the instructor can use the application to analyse her/his students' coursework results to identify the list of students who are in risk status that is likely to fail the final exam. These students strongly need support from the instructors to work hard to pass the course. The instructors can identify those students and provide additional support like remedial classes to them. This is to improve the students' learning process.

The instructors can use the application to count the Pass and Failures in each assessment. The application uses the values of each counts to identify certain patterns like many failures, 100% passing rate, and so on.

The count of the pass and failures helps the instructors to identify the level of the students they have in their class. Before the completion of the course, they can change their teaching methods or strategies to accommodate the learning capabilities of those students. This will improve the instructor's teaching process and also the students' learning process. As mentioned before, teaching and learning are the vital element of an educational institution.

### 3.3 Decision Tree

In a decision tree, each branch node represents a choice between a number of alternatives and the leaf nodes represents the decision. Decision trees are used to gain information for decision-making process [2]. Decision tree starts with the root node. From this node, the subsequent nodes were generated recursively based in the feature selection technique.The decision tree induction algorithm ID3 uses entropy and information gain technique for attribute selection. The successor of ID3 is C4.5 which uses gain ratio as the technique for attribute selection. The bias produced by the information gain is removed by gain ratio. The feature/attribute selection techniquesnamely information gain and gain ratio are explained in the next section.

### 3.4 Feature Selection

According to Ian H. Witten and et al. [3], the selection of the root node or attribute for further split in the decision tree is based on the attribute which has a pure subset. A pure subset, which is also called as homogeneous, of an attribute will contain one set of values either "yes" or "no" and that will not be split further.Entropy is used to measure the purity of an attribute [8]. Entropy is measured in units called bits which is less than one. When the number of either "yes" or "no" is zero, the entropy will be zero. When the number of "yes" and "no" is equal, the entropy will be the maximum value which is one. The leaf node of the decision tree will be the pure subset. The reason for the minus sign in the Entropy formula is that the logarithms of the fractions are negative [3]. The formula to calculate Entrophy is given below [10]:

$$Split\ Entropy(S, A) = -\sum_{v \in values(A)} \frac{|S_v|}{|S|} \log_2 \frac{|S_v|}{|S|} \quad (1)$$

The dataset is split on the different attributes and the entropy for each branch is calculated and added to get the entropy for the split. This resulting entropy is subtracted from the entropy before the split. This resulting value is called information gain [8]. The formulas to calculate Information Gain is given below [10]:

$$Gain\ (S, A) = Entropy(S) - \sum_{v \in values(A)} \frac{|S_v|}{|S|} \log_2 \frac{|S_v|}{|S|} \quad (2)$$

The information gain value for certain attributes like ID Number will give highest value. This value will be greater than the information gain of any other attributes, so ID Number will inevitably be chosen as the splitting attribute. But branching on the ID Number will not tell anything about the structure of the decision. To overcome this problem, gain ratio is used [3]. The formula to calculate Gain Ratio is given below [10]:

$$Gain\ Ratio = \frac{Gain\ (S,A)}{Split\ Entrophy\ (S,A)} \quad (3)$$

### 4. EXPERIMENTAL RESULTS

Sample of 20 students' data were obtained from Jubail University College, Kingdom of Saudi Arabia. The training data set is given in Table 2. The attributes selected to analyse the students' performance are Quiz 1, Quiz 2, Mid-Term, Assignment 1 and Assignment 2.

**TABLE 2: TRAINING DATA SET**

| S. No. | Quiz 1 | Quiz 2 | Mid-Term | Assignment 1 | Assignment 2 | Final |
|---|---|---|---|---|---|---|
| 1 | Pass | Pass | Pass | Pass | Pass | Pass |
| 2 | Pass | Pass | Pass | Pass | Pass | Pass |
| 3 | Fail | Pass | Fail | Pass | Pass | Fail |
| 4 | Pass | Pass | Pass | Pass | Pass | Pass |
| 5 | Fail | Fail | Fail | Pass | Fail | Fail |
| 6 | Fail | Fail | Pass | Pass | Pass | Fail |
| 7 | Pass | Fail | Pass | Pass | Pass | Fail |
| 8 | Pass | Pass | Pass | Pass | Pass | Pass |
| 9 | Pass | Pass | Pass | Pass | Pass | Pass |
| 10 | Fail | Fail | Pass | Pass | Pass | Fail |
| 11 | Pass | Pass | Pass | Pass | Pass | Pass |
| 12 | Pass | Pass | Fail | Pass | Pass | Pass |
| 13 | Pass | Pass | Pass | Pass | Pass | Pass |
| 14 | Pass | Pass | Pass | Pass | Pass | Pass |
| 15 | Pass | Fail | Pass | Pass | Pass | Fail |
| 16 | Pass | Pass | Pass | Pass | Pass | Pass |
| 17 | Pass | Pass | Pass | Pass | Pass | Pass |
| 18 | Pass | Fail | Fail | Pass | Fail | Fail |
| 19 | Pass | Pass | Pass | Pass | Pass | Pass |
| 20 | Pass | Pass | Pass | Pass | Pass | Pass |

First, the Entropy of S is calculated, where S is a set of 13 Pass and 7 Fail values in 20 students' records.

$$Entropy(S) = -\left(\frac{13}{20} \log_2 \frac{13}{20} + \frac{7}{20} \log_2 \frac{7}{20}\right)$$

$$Entropy(S) = 0.9315$$

In order to find the best attribute for a particular node in the decision tree we use the measure called gain ratio. To calculate the gain ratio, we need to calculate the Information Gain and Split Entropy. The information gain values for the five attributes such as Quiz 1, Quiz 2, Assignment 1, Assignment 2 and Mid-Term are given in Table 3.

**TABLE 3: INFORMATION GAIN VALUES OF ATTRIBUTES**

| Information Gain | Value |
|---|---|
| Information Gain(S, Quiz 1) | 0.7365 |
| Information Gain(S, Quiz 2) | 0.8645 |
| Information Gain(S, Assignment 1) | 0.5275 |
| Information Gain(S, Assignment 1) | 0.6265 |
| Information Gain(S, Mid-Term) | 0.5825 |

The split entropy values for the five attributes such as Quiz 1, Quiz 2, Assignment 1, Assignment 2 and Mid-Term are given in Table 4.

**TABLE 4: SPLIT ENTROPY VALUES OF ATTRIBUTES**

| Split Entropy | Value |
|---|---|
| Split Entropy(S, Quiz 1) | 0.195 |
| Split Entropy (S, Quiz 2) | 0.067 |
| Split Entropy (S, Assignment 1) | 0.404 |
| Split Entropy (S, Assignment 1) | 0.305 |
| Split Entropy (S, Mid-Term) | 0.349 |

The gain ratio values for the five attributes such as Quiz 1, Quiz 2, Assignment 1, Assignment 2 and Mid-Term are given in Table 5.

**TABLE 5: GAIN RATIO VALUES OF ATTRIBUTES**

| Gain Ratio | Value |
|---|---|
| Gain Ratio(S, Quiz 1) | 3.78 |
| Gain Ratio (S, Quiz 2) | 12.90 |
| Gain Ratio (S, Assignment 1) | 1.31 |
| Gain Ratio (S, Assignment 1) | 2.05 |
| Gain Ratio (S, Mid-Term) | 1.67 |

The decision tree is constructed using the gain ratio which is given above is given in Figure 1.

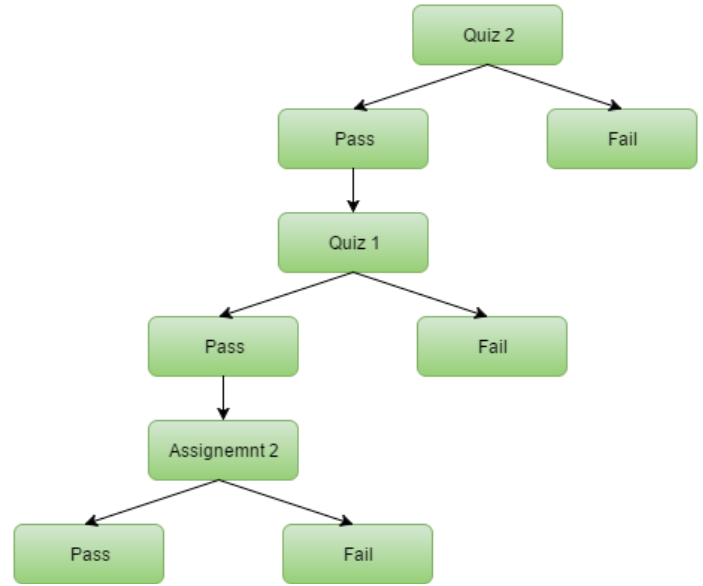

**FIGURE 1: DECISION TREE**

The decision tree given above is represented as IF-THEN rules and is given below:

IF Quiz 2 = "Pass" AND Quiz 1 = "Pass" AND Assignment 2 = "Pass" THEN Final = "Pass"
IF Quiz 2 = "Fail" OR Quiz 1 = "Fail" OR Assignment 2 = "Fail" THEN Final = "Fail"

The three attributes such as Quiz 1, Quiz 2 and Assignment 2 determines the students who are at risk which means will fail the final exam. The other two attributes such as Assignment 1 and Mid-Term were not used to determine the students at risk. Even though a student passes the Assignment 1 or Mid-Term and failed in any of the three assessment such as Quiz 1 or Quiz 2 or Assignment 2, he/she surely will be at risk of failing the final exam. For example, student number 7 in Table 2 passed the Mid-Term, Assignment 1, Assignment 2, and Quiz 1 but failed Quiz 2 and that student failed the final exam. As a conclusion, the students at risk are determined by the three attributes such as Quiz 1, Quiz 2, and Assignment 2.

## 5. CONCLUSION

In this paper, decision tree of Data Mining Classification technique was used on the students' assessment outcome database to identify the students who need support to perform well in the final examination to pass the course. This will also help the teachers / instructors to find out how well their current teaching strategy suits their present students' learning capabilities. This will help the students and the teachers to improve their learning and teaching processes respectively. As for further study, the analysis of students' data can be enhanced to a broad level to enhance the student retention rate.

Ms. Ganesan Kavitha received her M.Sc. in Computer Science from Bharathidasan University, India and M.Phil. in Computer Science from Mother Terasa University for Women, India. She worked as a Lecturer in The Thassim Beevi Abdul Kader College for Women, India for 4+ years and in INTI College Malaysia, Malaysia for 4+ years. Currently, she is working as a Lecturer in Jubail University College, Saudi Arabia.

Dr. Lawrance Raj received his M.Phil. from Manonmaniam Sundaranar University, India and PhD from Vinayaka Missions University. He is the Director of Department of MCA at Ayya Nada Janaki Ammal College in India from 2011 till date.